\documentstyle[12pt]{article} \textwidth=17,5cm\textheight=24,5cm
\begin{document} \title{True energy-momentum tensors are unique.\\
Electrodynamics spin tensor is not zero}
\author {Radi I. Khrapko\thanks{Email address:
khrapko\_ri@mai.ru}
\\ Moscow Aviation Institute, 125993, Moscow, Russia} \date{} \maketitle
\begin{abstract}

A true energy-momentum tensor is unique and does not admit an addition
of a term. The true electrodynamics' energy-momentum tensor is the
Maxwell-Minkowski tensor. It cannot be got with the Lagrange formalism.
The canonical energy-momentum and spin tensors are out of all relation to 
the
physical reality. The true electrodynamics' spin tensor is not equal to a 
zero. So,
electrodynamics' ponderomotive action comprises a force from the Maxwell
stress tensor and a torque from the spin tensor.

A gauge non-invariant expression for the spin tensor is presented and is
used for a consideration of a circularly polarized standing wave.
A circularly polarized light beam carries a spin angular momentum and an
orbital angular momentum. So, we double the beam's angular momentum. The 
Beth's
experiment is considered.


\end{abstract}

\section{An outlook on the standard electrodynamics}
The main points in the electrodynamics are an electric current four-vector
density $j^\alpha$ and electromagnetic field which is described by a
covariant skew-symmetric electromagnetic field tensor $F_{\mu\nu}$, or
by a contravariant tensor $F^{\alpha\beta}=
F_{\mu\nu}g^{\alpha\mu}g^{\beta\nu}$ ($\alpha,\:\nu,\ldots=0,\:1,\:2,\:3$,
or $=t,\:x,\:y,\:z$). The four-vector density $j^\alpha$ comprises a charge
density
and an electric current three-vector density: $j^\alpha=\{j^0=\rho,\:j^i\}$
($i,\:k,\ldots=1,\:2,\:3$, or $=x,\:y,\:z$).
$g^{\alpha\mu}=\{g^{00}=1,\:g^{ij}=-\delta^{ij}\}$.

We say that $ j^0$ and $ j^i$ are {\em coordinates} of the vector
$j^\alpha$. We write coordinates in braces.

Instead of $F_{\mu\nu}$ and $F^{\alpha\beta}$ we use
$$B_{\mu\nu}\stackrel{def}{=}F_{\mu\nu}\quad{\rm and}\quad
H^{\alpha\beta}\stackrel{def}{=}F^{\alpha\beta}.$$
$B_{\mu\nu}$ comprises electric strength and magnetic induction;
$H^{\alpha\beta}$ comprises electric induction and magnetic strength
$$ B_{\mu\nu}=\{B_{i0}=E_i,\:B_{ij}\};\qquad
H^{\alpha\beta}=\{H^{0i}=D^i,\:H^{ij}\}$$

An interaction of the electric current and the electromagnetic field results
in a four-force density
$f_\mu=j^\nu B_{\mu\nu}$ comprising a power density and a Lorentz force
density,
$$f_\mu=j^\nu B_{\mu\nu}=
\{p=-j^kE_k,\:f_i=\rho E_i+j^kB_{ik}\}$$
This force acts on the field. $-j^\nu B_{\mu\nu}$ acts on the current.

The second pair of Maxwell equations relates
the contravariant electromagnetic tensor to the electric current,
$$j^\alpha=\partial_\beta H^{\alpha\beta},$$
while covariant electromagnetic tensor, which is a differential form, is
closed,
$$\partial_{[\sigma} B_{\mu\nu]}=0.$$
It is the first pair of Maxwell equations. The first pair is a consequence 
of
an absence of a magnetic current $\xi_{\sigma\mu\nu}$,
$$3\partial_{[\sigma} B_{\mu\nu]}= \xi_{\sigma\mu\nu}=0.$$
If the magnetic current is absent, the covariant electromagnetic tensor can
be written as an exterior derivative of a covariant four-vector magnetic
potential ${\cal A}_\nu$,
$$B_{\mu\nu}=2\partial_{[\mu}{\cal A}_{\nu]}.$$
The four-vector ${\cal A}_\nu$ comprises a scalar potential and a covariant
three-vector magnetic potential,:
$${\cal A}_\nu=\{{\cal A}_0=\phi,\: {\cal A}_j=A_j\}$$
So, ${\cal A}^i=-A^i$.

\section{Energy-momentum tensor}
The four-force density $f_\mu=j^\nu B_{\mu\nu}$ can be written as
a divergence of a tensor $T^\alpha_\mu$ (more precisely, tensor density),
$$f_\mu=j^\nu B_{\mu\nu}=\partial_\alpha T^\alpha_\mu.$$
The tensor $T^\alpha_\mu$ is known as an energy-momentum tensor.
Maxwell and Minkowski found this tensor from experimental data:
$$T^\alpha_\mu=-B_{\mu\nu}H^{\alpha\nu}+ \delta^\alpha_\mu
B_{\sigma\nu}H^{\sigma\nu}/4.$$
The contravariant Maxwell-Minkowski tensor is symmetric,
$$T^\alpha_\mu g^{\mu\beta}= T^{\beta\alpha}= T^{(\beta\alpha)},\quad
T^{[\beta\alpha]}=0.$$

Now let us calculate a four-momentum which a field gets within a four-volume
$\Omega$,
$$P_\mu=\int_\Omega f_\mu d\Omega= \int_\Omega\partial_\alpha
T^\alpha_\mu d\Omega.$$
By the Stokes theorem, we can write this integral as a surface integral over
the supersurface of the four-volume,
$$P_\mu= \int_\Omega\partial_\alpha T^\alpha_\mu d\Omega=
\oint_{\partial\Omega} T^\alpha_\mu dV_\alpha,$$
where $dV_\alpha$ is a three-element of the supersurface. So, if a field is
bounded locally by an infinitesimal element $dV_\alpha$, the element gets
the infinitesimal four-momentum
$$dP_\mu= T^\alpha_\mu dV_\alpha.$$
This equation may be interpreted as a definition of an energy-momentum 
tensor.
\cite{GR7,GR10}

Let $dV_\alpha$ is timelike, i.e. it contains the time axis. Let, for 
example,
$dV_\alpha= \{dV_0=0,\:dV_i=da_i dt\}$, where $da_i$ is a two-element
of a surface. Then a four-force
$$dF_\mu=dP_\mu/dt= T^i_\mu da_i= \{d{\cal P}=T^i_0 da_i,\:
dF_j=T^i_jda_i\}$$
comprises a power $dP$ and a force $dF_j$ which are associated with the
element $da_i$. So,
$$T^i_0= {\cal S}^i= -B_{0j}H^{ij}=E_j\epsilon^{ijk}H_k=
\hbox{\bf E}\times\hbox{\bf H}$$
is the Poynting's vector, and $T^i_j$ is the Maxwell stress tensor.

If $dV_\alpha$ is spacelike, $dV_\alpha= \{dV_0=dV,\:dV_i=0\}$, then a
four-momentum
$$dP^\mu =T^{\mu0}dV_0 =\{d{\cal E} =T^{00}dV,\:dP^j =T^{j0}dV\}$$
is associated with the volume element dV. So,
$$T^{j0} =-g^{ji}B_{ik}H^{0k} =\delta^{ji}\epsilon_{ikn}B^nD^k
=\hbox{\bf D}\times\hbox{\bf B}$$
is the momentum density.

\section{Canonical energy-momentum tensor}
Apparently, the only theoretical way of getting Maxwell-Minkowski tensor
is a variation of the canonical Lagrangian
$$\mathop\Lambda\limits_c= -\frac{1}{4}B_{\mu\nu}H^{\mu\nu}\sqrt{-
g}$$
with respect to the metric tensor in the Minkowski space~
\cite[Sec.~94]{Landau}. But we do not consider this way here. Within the
scope of the standard Lagrange formalism the canonical Lagrangian gives a
canonical energy-momentum tensor \cite[Sec. 33]{Landau},
$$\mathop T\limits_c{}^\alpha_\mu= -H^{\alpha\nu}\partial_\mu {\cal
A}_\nu+\delta^\alpha_\mu B_{\nu\sigma}H^{\nu\sigma}/4.$$

This tensor is ``conserved'', i.e.
$$\partial_\alpha\mathop T\limits_c{}^\alpha_\mu =0,$$
due to the uniformity of space-time according to the Noether theorem. But
this tensor is out of all relation to the physical reality. It contradicts
experience \cite{GR10}.
For example, in a constant uniform $x$ directed magnetic field,
$B^x = B, B^y = B^z = 0, $
$$B_{yz}=H^{yz} =B, \quad {\cal A}_y= -Bz/2, \quad {\cal A}_z= By/2, $$
the canonical tensor gives zero value of a field pressure across field 
lines:
$${\mathop T\limits_c}{}^y_y ={\mathop T\limits_c}{}^z_z = 0,$$
what is wrong. Besides, the divergence of the
canonical tensor is equal to a wrong expression
$$\partial_\alpha{\mathop T\limits_c}{}^\alpha_\mu =j^\nu\partial_\mu
{\cal A}_\nu.$$
Besides that, the contravariant canonical tensor is nonsymmetric,
$${\mathop T\limits_c}^{[\alpha\beta]}= -\partial_\mu({\cal
A}^{[\alpha}\partial^{\beta]}{\cal A}^\mu).$$

To turn the canonical energy-momentum tensor to Maxwell-Minkowski
tensor theorists simply add two {\em ad hoc} terms
\cite[Sec.~33]{Landau}:
$$\mathop T\limits_c{}^\alpha_\mu + \partial_\nu({\cal A}_\mu
H^{\alpha\nu})- {\cal A}_\mu j^\alpha= T^\alpha_\mu.$$
The second term repairs the divergence of the tensor, and than the first 
term
symmetrizes the contravariant form of the canonical tensor. Certainly, this
addition has no basis and is completely arbitrary. So, we have to realize 
that
Lagrange formalism does not give a true energy-momentum tensor.

\section{The uniqueness of a true energy-momentum tensor}
Theorists simply ignore the second term of the addition, $-{\cal A}_\mu
j^\alpha$. They do not see it. But they attempt to explain an addition of
the first term, $\partial_\nu({\cal A}_\mu H^{\alpha\nu})$. For example,
Landau and Lifshitz wrote \cite[Sec.~32]{Landau},

``It is necessary to point out that the definition of the (energy-momentum)
tensor $T^{\alpha\beta}$ is not unique.
In fact, if ${\mathop T\limits_c}^{\alpha\beta}$ is defined by
$$\mathop T\limits_c{}^\alpha_\mu =
q_{,\mu}\cdot\frac{\partial\Lambda}{\partial q_{,\alpha}}-
\delta^\alpha_\mu\Lambda,\eqno(32.3)$$
then any other tensor of the form
$$\mathop T\limits_c{}^{\alpha\beta} + \frac{\partial}{\partial
x^\gamma}\psi^{\alpha\beta\gamma}, \qquad \psi^{\alpha\beta\gamma} = -
\psi^{\alpha\gamma\beta} \eqno(32.7)$$
will also satisfy equation
$$\partial T^{\alpha\beta}/\partial x^\beta = 0, \eqno(32.4)$$
since we have identically $\partial^{\rm
2}\psi^{\alpha\beta\gamma}/\partial x^\beta\partial x^\gamma = 0$. The
total four-momentum of the system does not change, since \dots we can
write
$$\int\frac{\partial\psi^{\alpha\beta\gamma}}{\partial{x}^\gamma}dV_
\beta = \frac{1}{2} \int\left(dV_\beta 
\frac{\partial\psi^{\alpha\beta\gamma}}
{\partial x^\gamma}-
dV_\gamma\frac{\partial\psi^{\alpha\beta\gamma}}{\partial
x^\beta}\right) =
\frac{1}{2}\oint\psi^{\alpha\beta\gamma}da_{\beta\gamma},$$
were the integral on the right side of the equation is extended over the
(ordinary) surface which `bond' the hypersurface over which the integration
on the left is taken. This surface is clearly located at infinity in the
three-dimensional space, and since neither field nor particles are present 
at
infinity this integral is zero. Thus the fore-momentum of the system is, as 
it
must be, a uniquely determined uantity.''

Exactly the same explanation was given by L. H.
Ryder~\cite[Sec.~3.2]{Ryder}.

But it seems to be incorrect \cite{GR7,GR10}.
$$\oint\psi^{\alpha\beta\gamma}da_{\beta\gamma} = 0$$
only if $\psi^{\alpha\beta\gamma}$ decreases on infinity rather quickly.
We present here a three-dimensional analogy concerning an electric current
$I$ and its magnetic field $H^{ij}$:
$$\frac{1}{2}\oint{H}^{ij}dl_{ij} = \int\partial_j{H}^{ij}da_i= \int j^i
da_i= I \neq 0.$$

So, the addition of $\partial_\gamma\psi^{\alpha\beta\gamma}$ can change
the total 4-momentum of a system. For example, it is easy to express the
energy-momentum tensor of an uniform ball of radius $R$ in the form
$\partial_\gamma \psi^{\alpha\beta\gamma}$:
$$\psi^{00i} = - \psi^{0i0} = \epsilon x^i/3 \quad(r< R), \quad \psi^{00i} =
- \psi^{0i0} = \epsilon R^{3}x^i/3r^{3} \quad (r > R)$$
$$\mbox{give }\quad\partial_i\psi^{00i}=\epsilon \quad (r<R),
\quad\partial_i\psi^{00i}=0 \quad (r > R).$$
Obviously, an addition of $\partial_\gamma\psi^{\alpha\beta\gamma}$
can change the total 4-momentum and changes the medium locally.

\section{Electrodynamics' spin tensor}
So, the Lagrange formalism does not give a true energy-momentum tensor.
But the formalism gives a more important thing. The formalism gives raise
an idea of a classical spin. A spin tensor $\mathop\Upsilon\limits_c
{}^{\mu\nu\alpha},$ coupled with the energy-momentum tensor, arises due
to the isotropy of space-time according to the Noether theorem. Essentially,
we have a pair of the canonical tensors
$$\mathop T\limits_c{}^{\mu\alpha}= -H^{\alpha\nu}\partial^\mu {\cal
A}_\nu+g^{\mu\alpha} B_{\nu\sigma}H^{\nu\sigma}/4, \quad
\mathop\Upsilon\limits_c{}^{\mu\nu\alpha} =-2 {\cal
A}^{[\mu}H^{\nu]\alpha}.$$

The existence of the spin tensor imply that electromagnetic field acts on 
its
boundary not only with the Maxwell stress tensor $T^{ji}$ but also with a
screw tensor $\Upsilon^{jki}$ (they are rather tensor densities). The stress
tensor provides a force acting on a surface element, and the screw tensor
provides a torque acting on the surface element,
$$dF^j= T^{ji}da_i, \qquad d\tau^{jk}= \Upsilon^{jki}da_i.\eqno (1)$$
In Minkowski space we have
$$dP^\mu= T^{\mu\alpha} dV_\alpha, \qquad
dS^{\mu\nu}=\Upsilon^{\mu\nu\alpha}dV_\alpha.$$
So, if a field is bounded locally by an infinitesimal element $dV_\alpha$,
the element gets the infinitesimal spin $dS^{\mu\nu}$.

It is natural that the canonical spin tensor is wrong just as the canonical
energy-momentum tensor is. Indeed, let us consider, for example, a $z$
directed circularly polarized plane wave:
$$\hbox{\bf E}=\hbox{\bf D}=(\hat{\hbox{\bf x}} -i\hat{\hbox{\bf y}})
e^{i\omega(t-z)}, \quad
\hbox{\bf H}=\hbox{\bf B} =\omega\hbox{\bf A} =i\hbox{\bf E}.$$
The coordinate
$$\mathop\Upsilon\limits_c{}^{zxy} = {\cal A}^xH^{zy}= A^xH_x$$
of the spin tensor is a current density of spin angular momentum about the
$y$ axis along the $y$ axis. This quantity is not zero, what is not rightly.

Nevertheless a density of the spin about the $z$ axis
$$\mathop\Upsilon\limits_c{}^{jk0} = -2{\cal A}^{[j}H^{k]0}=-
2A^{[j}D^{k]}= \hbox{\bf D}\times\hbox{\bf A},$$
and a current density of the spin about the $z$ axis along the $z$ axis
$$\mathop\Upsilon\limits_c{}^{jkz} = -2{\cal A}^{[j}H^{k]z}= \hbox{\bf
A}\cdot\hbox{\bf H}$$
correspond to reality for a plane wave (see Sec. 7).

Here a problem arises: what is an electrodynamics' true spin tensor. What
must we add to the canonical spin tensor to get the true spin tensor?

Our answer is as follows: a spin addition,
$\Delta\mathop\Upsilon\limits_c{}^{\mu\nu\alpha}$, and the
energy-momentum addition,
$$\Delta\mathop T\limits_c{}^{\mu\alpha} = \partial_\nu({\cal A}^\mu
H^{\alpha\nu})- {\cal A}^\mu j^\alpha,$$
must satisfy the equation,
$$\partial_\alpha(\Delta\mathop\Upsilon\limits_c{}^{\mu\nu\alpha}) =
2\Delta\mathop T\limits_c{}^{[\mu\nu]}.\eqno (2)$$

It is easy to find from (2) that
$$\Delta\mathop\Upsilon\limits_c{}^{\mu\nu\alpha}= 2{\cal
A}^{[\mu}H^{\nu]\alpha} + 2{\cal A}^{[\mu}
\partial^{\mid\alpha\mid}{\cal A}^{\nu]},$$
and so we obtain \cite{GR10}
$$\Upsilon^{\mu\nu\alpha} =\mathop\Upsilon\limits_c{}^{\mu\nu\alpha}
+\Delta\mathop\Upsilon\limits_c{}^{\mu\nu\alpha} = 2{\cal
A}^{[\mu}\partial^{\mid\alpha\mid}{\cal A}^{\nu]}.$$

Theorists recognize the equation (2), but they consider only the first term 
of
$\Delta\mathop T\limits_c{}^{\mu\alpha}$, that is $\partial_\nu({\cal A}^\mu
H^{\alpha\nu}),$ as the energy-momentum addition. As a result, the
equation (2) gives
$$\Delta\tilde{\mathop\Upsilon\limits_c}{}^{\mu\nu\alpha}=
2 {\cal A}^{[\mu}H^{\nu]\alpha}, \quad {\rm and} \quad
\tilde\Upsilon{}^{\mu\nu\alpha} = 0.$$
That is why a classical spin is absent in the modern electrodynamics.That is
why they consider that a circularly polarized plane wave has no angular
momentum.

So, our true spin tensor
$$\Upsilon^{\mu\nu\alpha} = 2{\cal
A}^{[\mu}\partial^{\mid\alpha\mid}{\cal A}^{\nu]}.$$
is a function of the vector potential ${\cal A}_\mu$ and is not gauge 
invariant.
We greet this fact. As is shown \cite{0105031}, ${\cal A}^\mu$ must satisfy
the Lorentz condition, $\partial_\mu {\cal A}^\mu=0$.

\section{The plane wave problem}
The elimination of the electrodynamics' spin tensor gives rise an opinion
that total angular momentum $J^{ik}$ is a moment of the linear momentum,
i.e. the total angular momentum is an orbital angular momentum
\cite{Heitler,Jackson,Sim,All99,All00}
$$dJ^{ik}= dL^{ik}= 2r^{[i}dP^{k]},\eqno (3)$$
where $dP^k$ is proportional to Poynting's vector. This implies that a
circularly polarized plane wave has no angular momentum directed along
the propagation direction of the wave
\cite{Shapochnikov,Humblet,Heitler,Crichton,Ohanian}, that only a
quasiplane wave of finite transverse extent, i.e. a beam, carries an angular
momentum whose direction is along the propagation direction. In accord
with (3) this angular momentum is provided by an outer region of the beam
because a falloff in intensity gives rise $E$ and $B$ fields which are
parallel to wave vector, and so the energy flow has components
perpendicular to the wave vector \cite{Yurchenko}. They name this
angular momentum {\em spin} \cite{Ohanian}. Within an inner region of
the beam $E$ and $B$ fields are perpendicular to the wave vector, and the
mass-energy flow is parallel to the wave vector \cite{Jackson}. So, there is
no angular momentum in the inner region \cite{Ohanian}.

To refute this paradigm I proposed a specific experiment in Oct., 1999. I
proposed to consider a two-element flat target comprising an inner disc and
a closely fitting outer annulus \cite{Khrapko}. According to standard
electrodynamics, the inner part of the target does not perceive a torque
when the target absorbs a circularly polarized wave. But it is clear that
really the disc does perceive a torque from the wave. The disc will be
twisted in contradiction with the paradigm.

Allen and Padgett \cite{Allen} agree with an incorrectness of the opinion 
that
any plane wave has no angular momentum. But they try to endow a
circularly polarized plane wave with angular momentum within the scope of
the standard electrodynamics. The authors have attempted to explain the
torque acting on the disc within the scope of the standard electrodynamics.
They wrote, ``Any form of aperture introduces an intensity gradient \dots
and a field component is induced in the propagation direction and so the
dilemma is potentially resolved.''

Alas! A small clearance between the inner disc and outer annulus does not
aperture a wave and does not induce a field component in the propagation
direction. The imaginary decomposition of the plane wave into three beams,
the inner beam, the annular beam, and the remainder, is not capable to
create longitudinal field components and, correspondingly, transverse
momentum and a torque acting on the disc. Maxwell stress tensor cannot
supply the disc with a torque.

To resolve the dilemma we must use the conception of classical
electrodynamics' spin which is described by the spin tensor
$\Upsilon^{\mu\nu\alpha}$. So, we must recognize that the standard
classical electrodynamics is not complete. Electrodynamics' spin tensor is
not zero \cite{0105031}, and ``ponderomotive forces'' acting on a
surface element $da_i$ consist of both, the force itself, and a torque (1).

So, the annulus of our target perceives the orbital angular momentum
$L={\cal E}/\omega$, the disc perceives a spin angular momentum
$S={\cal E}/\omega$, and the target as a whole perceives a total angular
momentum
$$J=L+S=2{\cal E}/\omega.$$
So, we double the beam's angular momentum.

This conclusion, naturally, must not conflict with the Beth's famous 
experiment
\cite{Beth}. And it is the case! It was found that Beth's
half-wave plate perceives only spin angular momentum. The orbital angular
momentum is eliminated by an interference of the passing and returning light
beams in the experiment. Indeed, let us start from the Jackson's expression 
for a
circularly polarized beam \cite{Jackson}.
$${\bf E}(x,y,z,t)
=\Re\left\{[\hat{\bf x}+i\hat{\bf y}+\hat{\bf z}(i\partial_x-\partial_y)]
E_0(x,y)e^{i(z-t)}\right\},$$
$${\bf B}(x,y,z,t)
=\Re\left\{[\hat{-i\bf x}+\hat{\bf y}+\hat{\bf z}(\partial_x+i\partial_y)]
E_0(x,y)e^{i(z-t)}\right\}.$$
Here $E_{\rm 0} (x, y)$ is the electric field of the beam. $E_{\rm 0} (x, y) 
=
{\rm Const}$ inside the beam, and $E_{\rm 0} (x, y) =0$ outside the beam.

The returning light beam may be got by changing the sign of $z$ and $y$.
Adding up the  passing and returning light beams we get interesting 
expressions,
$${\bf E}_{\rm tot}
=2[E_0(\hat{\bf x}\cos z-\hat{\bf y}\sin z)
-\hat{\bf z}(\sin z\partial_x E_0+\cos z\partial_y E_0)]\cos t,$$
$${\bf B}_{\rm tot}
=-2[E_0(\hat{\bf x}\cos z-\hat{\bf y}\sin z)
-\hat{\bf z}(\sin z\partial_x E_0+\cos z\partial_y E_0)]\sin t.$$
The $E$ and $B$ fields are parallel everywhere. So, the Poynting vector is a
zero.

\section{The Humblet transformation}
A considerable amount of papers is devoted to a circularly polarized light
beam. They try to prove that the angular momentum (3) which is localized
at the surface of the beam is distributed over the body of the beam and
represents the spin of the beam.

A calculation of the orbital angular momentum (3),
$$L^{ij} =2\int r^{[i}T^{j]0}dV, $$
for the Jackson's expression gives
$$L^z =\int E^{\rm 2}_{\rm 0}dV/\omega, \qquad L^x = L^y=0.$$

The energy of the field,
$${\cal E}= \int T^{00}dV= \int E^{\rm 2}_{\rm 0}dV = \omega L^z,$$
and the ratio ${\cal E}/L= \omega$ is the same as the ratio ${\cal E}/S,$ 
i.e.
energy/spin, for a photon. So, $L=S$ for the beam. But it
does not follow that $L$ {\em is} $S.$ Simply, the total angular momentum
of the beam, $J,$ is twice the orbital angular momentum,
$$J^{ij}=L^{ij}+S^{ij}= 2\int r^{[i}T^{j]0}dV+ \int\Upsilon^{ij0}dV.$$

Another method for the calculation of the orbital angular momentum of the
beam was given by Humblet \cite{Humblet,Ohanian}.
He transforms angular momentum (3) into an integral of the coordinate
$\mathop\Upsilon\limits_c{}^{ij0},$
$$\hbox{\bf L}= \int\hbox{\bf r}\times (\hbox{\bf D}\times\hbox{\bf
B})dV \longrightarrow
\int\ (\hbox{\bf D}\times\hbox{\bf A})dV.$$

Let us consider this transformation.
\begin{eqnarray*}
L^{ij}&=&2\int r^{[i}T^{j]0}dV= -2\int r^{[i}g^{j]k}B_{kl}H^{0l}dV= -
4\int r^{[i}g^{j]k}\partial_{[k}{\cal A}_{l]}\cdot D^ldV\\
&=&-2\int r^{[i}\partial^{j]}{\cal A}_l\cdot D^ldV +2\int
r^{[i}\partial_l({\cal A}^{j]}D^l)dV.
\end{eqnarray*}
Ohanian \cite{Ohanian} writes, ``The first term in the Eq.
represents the orbital angular momentum, and the second term the spin.''
But the derivative $\partial^j{\cal A}_l$ has only $j=z$ coordinate inside
the beam, and, in any case, the term $\partial^j {\cal A}_l\cdot D^l$ is $z$
directed. So, the first term is an integral moment of a longitudinal
component of the momentum $ T^{j0}dV$ and is equal to a zero if the
origin of $r^i$ is at the axis of symmetry. The first term bears no relation
to the angular momentum of the beam.

The second term is transformed by an integration by parts,
$$2\int r^{[i}\partial_l ({\cal A}^{j]}D^l)dV= -2\int D^{[i}{\cal
A}^{j]}dV =\int (\hbox{\bf D}\times\hbox{\bf A})dV.$$
And again we have $L=S$, but not $L$ is $S$.

{\em Vice versa}. Since for any compact tensor field $I^{ij}$
$$\int I^{[ij]}dV =\int r^{[i}\partial_kI^{j]k}dV,$$
we have if
$$I^{ij}=\mathop\Upsilon\limits_c{}^{ij0}=2{\cal A}^{[i}D^{j]},\quad
S^{ij}=\int\mathop\Upsilon\limits_c{}^{ij0}dV
=+2\int r^{[i}\partial_k ({\cal A}^{j]}D^k)dV
=2\int r^{[i}T^{j]0}dV.$$

Ohanian \cite{Ohanian} pay attention to a mathematical
equivalence between the equation and calculating a magnetic moment,
$P^{ij}_{\rm m},$ of a body as a moment of the Amperian magnetization
current $j^i,$
$$P^{ij}_{\rm m}= \int I^{[ij]}dV =\int r^{[i}\partial_kI^{j]k}dV =\int
r^{[i}j^{j]}dV,$$
where $I^{ij}$ is the magnetization tensor, and $j^i=\partial_k I^{ik}.$

I am presenting another similar equation involving a magnetic strength
$H^{ij}$ and a current $j^i$,
$$\int H^{ij}dV =\int r^{[i}\partial_kH^{j]k}dV =\int r^{[i}j^{j]}dV,$$

But I think these relationships do not prove that a moment of a current and 
a
magnetic strength, a moment of a current and a magnetization, a moment of
a momentum and a spin have identical natures.

\section {Circularly polarized standing wave}
The electrodynamics is asymmetric. Magnetic induction is closed, but
magnetic field strength has electric current as a source:
$$\partial_{[\alpha}B_{\beta\gamma]}=0,\quad \partial_\nu H^{\mu\nu}
=j^\mu.$$
So, a magnetic vector potential exists, but, generally speaking, an electric
vector potential does not exist. However, when currents are absent the
symmetry is restored, and a possibility to introduce an electric multivector
potential $\Pi^{\mu\nu\sigma}$ appears. The electric multivector potential
satisfies the equation
$$\partial_\sigma\Pi^{\mu\nu\sigma}=H^{\mu\nu}.$$

A covariant vector, dual relative to the three-vector potential,
$$\Pi_\alpha=\epsilon_{\alpha\mu\nu\sigma}\Pi^{\mu\nu\sigma},$$
is an analog of magnetic vector potential. We name it electric vector
potential. In our case it satisfies the relationships:
$$\Pi_0=0,\quad\partial_0\Pi_i=-H_i,\quad H_i=\epsilon_{ijk}H^{jk}.$$

The symmetry of the electrodynamics forces us to offer a symmetric
expression for the spin tensor consisting of two parts, electric and 
magnetic.
$$\Upsilon^{\mu\nu\alpha} =
{\mathop\Upsilon\limits_e}^{\mu\nu\alpha}+
{\mathop\Upsilon\limits_m}^{\mu\nu\alpha}=
A^{[\mu}\partial^{\mid\alpha\mid}A^{\nu]}
+\Pi^{[\mu}\partial^{\mid\alpha\mid} \Pi^{\nu]}.$$

Here the symmetric expression is applied to a circularly polarized standing
wave. We consider such a wave which falls upon a superconducting 
$x,y$-plain,
and is reflected from it. The energy current density is equal to a zero
in the wave, $T^{tz} = \mbox{\bf E}\times \mbox{\bf H} = 0$. But the
electrical and magnetic energy densities vary with $z$ in anti-phase. So the
total energy density is constant. The momentum current density, i.e. the
pressure, is constant too:
$$E^{\rm 2}/2=1-\cos2kz, \quad H^{\rm 2}/2=1+\cos2kz, \quad
T^{tt}=T^{zz}=(E^{\rm 2}+H^{\rm 2})/2=2.$$

The spin current density must be zero, $\Upsilon^{xyz}= 0$, and it is
expected that the spin density comprise electrical and magnetic parts which
are shifted relative to one another. This result is obtained below.

A circularly polarized plane wave which propagates along $z$-direction
involves the vectors {\bf H}, {\bf E}, {\bf A}, $\Pi$ which lay in 
$xy$-plane,
and we shall represent them by complex numbers instead of real parts
of complex vectors. For example,
$$\mbox{\bf H}=\{H^x,\ H^y\}\to\ H=H^x+iH^y.$$
Then the product of a complex conjugate number $\overline{E}$ and other
number $H$ is expressed in terms of scalar and vector products of the
corresponding vectors. For example,
$$\overline{E}\cdot H=
(\mbox{\bf E}\cdot\mbox{\bf H})+i(\mbox{\bf E}\times\mbox{\bf
H})^z.$$

Since all this vectors do not vary with $x$ and $y$, then
$${\rm curl}\mbox{\bf H}=
\{-\partial_zH^y,\ \partial_zH^x\}\to i\partial_zH,\quad
{\rm curl}^{-1}\to -i\int dz.$$

The angular velocity of all the vectors is $\omega$ and the wave number
along $z$-axis is $k =\omega.$ Therefore
$$\mbox{\bf H}\to H_{01}e^{i\omega(t-z)}\
\mbox{or, for a reflected wave, } H_{02}e^{i\omega(t+z)},$$
$$\partial_t\to i\omega,\quad\partial_z\to\mp i\omega,\quad
{\rm curl}\to\pm\omega,\quad{\rm curl}^{-1}\to\pm1/\omega.$$

If $z = 0$ at the superconducting $x,y$-plain, then the falling and
reflected waves are recorded as
$$H_1=e^{i\omega(t-z)},\quad E_1=-ie^{i\omega(t-z)},\quad
H_2=e^{i\omega(t+z)},\quad E_2=ie^{i\omega(t+z)}.$$
The complex amplitudes are equal here:
$H_{01}=H_{02}=1,\ E_{01}=-i,\ E_{02}=i.$

Since
$\mbox{\bf A}={\rm curl}^{-1}\mbox{\bf H},\
\Pi={\rm curl}^{-1}\mbox{\bf E},$
the other complex amplitudes are received by a simple calculation (time
derivative is designated by a point):
$$A_{01}=1/\omega,\, \dot A_{01}=i,\,
\Pi_{01}=-i/\omega,\, \dot\Pi_{01}=1,\,
A_{02}=-1/\omega,\, \dot A_{02}=-i,\,
\Pi_{02}=-i/\omega,\, \dot\Pi_{02}=1.$$

Now we calculate the electric and magnetic parts of the volumetric spin
density.
\begin{eqnarray*}
{\mathop\Upsilon\limits_e}^{xyt}&=&
(\mbox{\bf A}\times\dot{\mbox{\bf A}})/2=
\Im(\overline{(A_1+A_2)}\cdot({\dot A}_1+{\dot A}_2))/2\\
&=&\Im((e^{-i\omega(t-z)}-e^{-i\omega(t+z)})i
(e^{i\omega(t-z)}-e^{i\omega(t+z)}))/2\omega=
(1-\cos 2\omega z)/\omega,
\end{eqnarray*}
$${\mathop\Upsilon\limits_m}^{xyt}=
(\Pi\times\dot\Pi)/2=
\Im(\overline{(\Pi_1+\Pi_2)}\cdot({\dot\Pi}_1+{\dot\Pi}_2))/2=
(1+\cos 2\omega z)/\omega,$$
$$\Upsilon^{xyt}=
{\mathop\Upsilon\limits_e}^{xyt}+
{\mathop\Upsilon\limits_m}^{xyt}=2/\omega.$$
So, the terms which oscillate along $z$-axis are canceled out.
It is easy to calculate that the spin current density is equal to a zero
(the prime denote the derivative with respect to z):
$${\mathop\Upsilon\limits_e}^{xyz}=
-(\mbox{\bf A}\times{\mbox{\bf A}}')/2=0, \quad
{\mathop\Upsilon\limits_m}^{xyz}=
-(\Pi\times\Pi')/2=0.$$

\section*{Notes}
The material of this paper was rejected or ignored by the journals (dates
of the submissions are in brackets):
Phys. Rev. D (25 Sep 2001), Foundation of Physics (28 May 2001),
American J. of Physics (15 Sep 1999, 10 Sep 2001, 28 Mar 2002), Acta
Physica Polonica B (28 Jan 2002, 09 May 2002), Phys. Lett. A (22 July 2002),
Experimental \& Theor. Phys. Lett. (14 May 1998, 17 June 2002), J.
Experimental \& Theor. Phys. (27 Jan 1999, 25 Feb 1999, 13 Apr 2000, 25 May
2000, 16 May 2001, 26 Nov 2001), Theor. Math. Phys. (29 Apr 1999, 17 Feb
2000, 29 May 2000, 18 Oct 2000), Physics - Uspekhi (25 Feb 1999, 12 Jan
2000, 31 May 2000), Rus. Phys. J. (18 May 1999, 15 Oct 1999, 1 March 2000,
25 May 2000, 31 May 2001, 24 Nov 2001).

The material of this paper was rejected by the arXiv (21 Jan 2002,
18 Feb 2002, 02 June 2002, 13 June 2002).

The subject matter of this paper had been published on the web site:
\\ http://www.mai.ru/projects/mai\_works

\section*{Acknowledgements}

I am deeply grateful to Professor Robert H. Romer for publishing my
question \cite{Khrapko}.

\end{document}